\newfam\msbfam
\font\twlmsb=msbm10 at 12pt
\font\eightmsb=msbm10 at 8pt
\font\sixmsb=msbm10 at 6pt
\textfont\msbfam=\twlmsb
\scriptfont\msbfam=\eightmsb
\scriptscriptfont\msbfam=\sixmsb
\def\cj{\fam\msbfam}

\def\C{{\cj C}}

\def\R{{\cj R}}

\def\Z{{\cj Z}}

\centerline{\bf AHARONOV-BOHM EFFECT, DIRAC MONOPOLE, AND BUNDLE THEORY}

\

\centerline{M. Socolovsky*}

\

\centerline{\it  Instituto de Ciencias Nucleares, Universidad Nacional Aut\'onoma de M\'exico}
\centerline{\it Circuito Exterior, Ciudad Universitaria, 04510, M\'exico D. F., M\'exico} 

\

{\bf Abstract.} {\it We discuss the Aharonov-Bohm ($A-B$) effect and the Dirac ($D$) monopole of magnetic charge $g={{1}\over{2}}$ in the context of bundle theory, which allows to exhibit a deep geometric relation between them. If $\xi_{A-B}$ and $\xi_D$ are the respective $U(1)$-bundles, we show that $\xi_{A-B}$ is isomorphic to the pull-back of $\xi_D$ induced by the inclusion of the corresponding base spaces $\iota:(D_0^2)^*\to S^2$. The fact that the $A-B$ effect disappears when the magnetic flux in the solenoid equals an integer times the quantum of flux $\Phi_0={{2\pi}\over{\vert e\vert}}$ associated with the electric charge $\vert e\vert$, reflects here as a consequence of the pull-back by $\iota$ of the Dirac connection in $\xi_D$ to $\xi_{A-B}$, and the Dirac quantization condition. We also show the necessary vanishing in $\xi_{A-B}$ of the pull-back of the Chern class $c_1$ in $\xi_D$.} 

\


\


\

{\bf 1. Introduction}

\

As is well known, the Aharonov-Bohm ($A-B$) effect [1] and the Dirac ($D$) magnetic monopole [2],[3] proposal have had a profound influence in the development of the gauge theories of fundamental interactions. The first one of these phenomena was immediately verified experimentally [4] and by many others later on [5], while even if Dirac monopoles have not yet being seen in Nature, both grand unified theories [6] and string theories [7] predict their existence.

\

The description of both the $A-B$ effect and the $D$ monopole are deeply rooted in the concept of gauge potential and therefore in the concept of connection in fiber bundles. The first one provides an explicit evidence of the non-local character of quantum mechanics describing the motion of electrically charged particles in a non-simply connected space [8], while the second one makes unavoidable the use of at least two charts on manifolds to define the gauge potential, leading to the necessity of a description in terms of a non-trivial bundle [9].

\

The close relationship between both phenomena consists in the facts that when the magnetic flux $\Phi_{A-B}$ is an integer multiple of the quantum of flux $\Phi_0={{2\pi}\over{\vert e\vert}}$ associated with the electric charge $\vert e\vert$, the $A-B$ effect vanishes, and when $\Phi_{A-B}$ also equals the magnetic flux of the monopole, $\Phi_D$, the Dirac quantization condition ($D.Q.C.$) follows. In this note we want to emphasize this relation at a perhaps deeper level, namely through the relationship between the fiber bundles $\xi_{A-B}$ (trivial) and $\xi_D$ (non-trivial) in which both phenomena occur. After some basic material in section {\bf 2.}, in section {\bf 3.} we exhibit the bundle morphism $\xi_{A-B}\to\xi_D$ induced by the inclusion $\iota$ between the corresponding base spaces, and in section {\bf 4.} we use $\iota$ to construct the pull-back bundle $\iota^*(\xi_D)$, which in turn is proved, in section {\bf 5.}, to be isomorphic to $\xi_{A-B}$ i.e. $$\xi_{A-B}\cong\iota^*(\xi_D).\eqno{(1)}$$ This is the main result of the present paper, since it exhibits a deep geometric relation between the $A-B$ effect and the magnetic monopole. Of course, the pull-back of the first Chern class $c_1$ of $\xi_D$, $\iota^*(c_1)$, vanishes in $\xi_{A-B}$, what is proved in section {\bf 6.} In section {\bf 7.} we show that the pull-back of the Dirac connection from $\xi_D$ to $\xi_{A-B}$ leads to the vanishing of the $A-B$ effect when the $D.Q.C.$ holds, thus setting on purely geometric grounds, one of the basic relations between $A-B$ and $D$. Section {\bf 8} is devoted to final comments.

\

We use the natural system of units $\hbar =c=1$.

\

{\bf 2. Basics}

\

In Ref. [8], the $U(1)$-bundle associated with the $A-B$ effect [1] with an infinitesimally thin and infinitely long solenoid was shown to be the {\it product} -and therefore trivial- {\it bundle} $$\xi_{A-B}:S^1\to (T_0^2)^*\buildrel{pr_1}\over\longrightarrow (D_0^2)^* \eqno{(2)}$$ where $S^1=U(1)=\{z\in\C, \ \vert z\vert=1\}$ is the structure group, $(D_0^2)^*$ is the punctured open disk in two dimensions, $(T_0^2)^*=(D_0^2)^*\times S^1$ is the open solid 2-torus minus a circle, and $pr_1$ is the projection in the first entry. One has the homeomorphisms $(D_0^2)^*\cong (\R^2)^*=\R^2\setminus \{0\}\cong\C^*=\C\setminus\{0\}$. The reason for (2) is that, in the above conditions, by symmetry reasons the space available to the electrically charged particles (``electrons") moving around the solenoid is $(\R^2)^*$ which is of the same homotopy type as the circle $S^1$. Then the set of isomorphism classes of $U(1)$-bundles over $(\R^2)^*$ consists of only one element [10]: the class of the product (trivial) bundle $(T_0^2)^*$. 

\

On the other hand, the fiber bundles associated with Dirac monopoles [2],[3] of magnetic charge $g=\# k$ with $k$ an integer and $\#$ a number depending on units, are the Hopf bundles [9],[11] $$\xi_D^{(k)}:S^1\to P^3_k\buildrel{\pi_k}\over\longrightarrow S^2 \eqno{(3)}$$ where $P^3_0=S^2\times S^1$ (the trivial bundle), $P^3_k\cong P^3_{-k}$, $S^2$ is the 2-sphere with $S^2\cong\R^2\cup\{\infty\}\cong\C\cup\{\infty\}$. In particular, we are interested in the case $k=1$ for which $P^3_1\cong S^3$: the 3-sphere given by $$S^3=\{(z_1,z_2)\in\C^2, \ \vert z_1\vert^2+\vert z_2\vert^2=1\},\eqno{(4)}$$ $\pi_3\equiv\pi$ is the Hopf map [12] $$\pi:S^3\to S^2, \ (z_1,z_2)\mapsto\pi(z_1,z_2)=\{\matrix{z_1/z_2, \ z_2\neq 0\cr \infty, \ z_2=0\cr} \ .\eqno{(5)}$$ We denote this {\it non-trivial} bundle $\xi_D$: $$\xi^{(1)}_D\equiv \xi_D:S^1\to S^3\buildrel{\pi}\over\longrightarrow S^2.\eqno{(6)}$$ The global connection on $\xi_D$ corresponding to $g={{1}\over{2}}$ ($\#={{1}\over{2}}$ and $k=1$) is the 1-form $\omega\in\Omega^1S^3\otimes u(1)$, with $u(1)=Lie(U(1))=i\R$, given by [13] $$\omega={{i}\over{2}}(d\chi+cos\theta d\varphi),\eqno{(7)}$$ where $\chi$, $\theta$ and $\varphi$ are the Euler angles in $S^2$ or $\R^3$ ($\theta\in [0,\pi]$ and $\chi,\varphi\in [0,2\pi)$). The differential of $\omega$ is the 2-form $$d\omega=-{{i}\over{2}}sin\theta d\theta\wedge d\varphi=-F\in\Omega^2S^3\otimes u(1)\eqno{(8)}$$ where $F$ is the field strength $$F=i\vert\vec{B}\vert sin\theta d\theta\wedge d\varphi\eqno{(9)}$$ with $$\vec{B}=({{1}\over{2}}){{\vec{r}}\over{r^3}}\eqno{(10)}$$ the magnetic field of the monopole in $\R^3\setminus \{0\}$ (see below). 

\

$\omega$ can be read from the squared length element on $S^3$: $$dl^2_{S^3}(\chi,\theta,\varphi)={{1}\over{4}}(d\theta^2+sin^2\theta d\varphi^2+(d\chi+cos\theta d\varphi)^2)\eqno{(11)}$$ which, in turn, can be obtained from the identification of $S^3$ with the group $SU(2)$ with elements $$\pmatrix{z_1 & z_2\cr -\bar{z}_2 & \bar{z}_1\cr}=\pmatrix{e^{{{i}\over{2}}(\varphi+\chi)}cos{{\theta}\over{2}} & e^{{{i}\over{2}}(\varphi-\chi)}sin{{\theta}\over{2}}\cr -e^{-{{i}\over{2}}(\varphi-\chi)}sin{{\theta}\over{2}} & e^{-{{i}\over{2}}(\varphi+\chi)}cos{{\theta}\over{2}}}.\eqno{(12)}$$

\

Covering $S^2$ with the open sets $U_+$ and $U_-$ respectively defined by $\theta\in[0,\pi)$ (the south pole $S$ excluded) and $\theta\in(0,\pi]$ (the north pole $N$ excluded), considering the pull-back of $\omega$ to $S^2\setminus\{N,S\}$ with the local sections $$s_N:U_+\setminus\{N\}\to S^3, \ s_N(\hat{n})=(cos{{\theta}\over{2}},sin{{\theta}\over{2}}e^{i\varphi}),\eqno{(13a)}$$ $$s_S:U_-\setminus\{S\}\to S^3, \ s_S(\hat{n})=(cos{{\theta}\over{2}}e^{i\varphi},sin{{\theta}\over{2}}),\eqno{(13b)}$$ with $\hat{n}=(sin\theta cos\varphi, sin\theta sin\varphi, cos\theta)$, using the inclusion $$j:S^3\to\R^4, \ j(z_1,z_2)=(x_1,x_2,x_3,x_4)$$ $$=(cos({{\varphi+\chi}\over{2}})cos{{\theta}\over{2}},sin({{\varphi+\chi}\over{2}})cos{{\theta}\over{2}},cos({{\varphi-\chi}\over{2}})sin{{\theta}\over{2}},sin({{\varphi-\chi}\over{2}})sin{{\theta}\over{2}}), \eqno{(14)}$$ and defining the 1-form $\tilde{\omega}\in\Omega^1\R^4\otimes u(1)$ through $$\tilde{\omega}=i(x^1dx^2-x^2dx^1-x^3dx^4+x^4dx^3),\eqno{(15)}$$ one can prove that $j^*(\tilde{\omega})=\omega$ and that $s_{N,S}^*(\omega)$ are the usual local 1-forms $A_\pm$ on $S^2$, namely $$A_+(\theta,\varphi)=s_N^*(\omega)(\theta,\varphi)=(j\circ s_N)^*(\tilde{\omega})(\theta,\varphi)=-{{i}\over{2}}(1-cos\theta)d\varphi, \eqno{(16a)}$$ $$A_-(\theta,\varphi)=s_S^*(\omega)(\theta,\varphi)=(j\circ s_S)^*(\tilde{\omega})(\theta,\varphi)={{i}\over{2}}(1+cos\theta)d\varphi. \eqno{(16b)}$$ The corresponding $u(1)$-valued 3-vector potentials are $$\vec{A}_+=-i{{1-cos\theta}\over{2rsin\theta}}\hat{\varphi}, \ \vec{A}_-=+i{{1+cos\theta}\over{2rsin\theta}}\hat{\varphi},\eqno{(17a)}$$ defined also at $\theta=0$ ($\vec{A}_+$) and $\theta=\pi$ ($\vec{A}_-$): $$\vec{A}_+(\theta=0)=\vec{A}_-(\theta=\pi)=\vec{0}\eqno{(17b)}$$ and on a 2-sphere of arbitrary radius $r>0$. Clearly, the rotor of $\vec{A}_+$ and $\vec{A}_-$ gives the magnetic field $\vec{B}$. 

\

The first Chern class of $\xi_D$ (taking $S^2$ with unit radius) is given by $$c_1(\xi_D)={{i}\over{2\pi}}[F]\eqno{(18)}$$ where $[F]$ is the cohomology class of $F$ in $H^2(S^2)$: cohomology of the 2-sphere in dimension 2. The integral of ${{i}\over{2\pi}}F$ over $S^2$ gives the first Chern number of $\xi_D$: $${{i}\over{2\pi}}\int_{S^2}F=1.\eqno{(19)}$$ This means that the magnetic charge is a measure of the topological non-triviality of the bundle $\xi_D$ i.e. of the space where it ``lives". In other words, the monopole charge is not a property of the gauge field $A_\pm$ itself, but of the $U(1)$-bundle on which the monopole is a connection. 

\

{\bf 3. Bundle morphism $\xi_{A-B}\to \xi_D$}

\

Using the homeomorphisms $(D_0^2)^*\cong \C^*$ and $S^2\cong\C\cup\{\infty\}$, it can be easily verified that ($\iota$,$\bar{\iota}$) given by $$\iota:\C^*\to\C\cup\{\infty\}, \ \iota(z)=z\eqno{(20)}$$ and $$\bar{\iota}:\C^*\times S^1\to S^3, \ \bar{\iota}(z,e^{i\varphi})={{(z,1)}\over{\vert\vert(z,1)\vert\vert}}e^{i\varphi}\eqno{(21)}$$ with $\vert\vert(z,1)\vert\vert=\sqrt{1+\vert z\vert^2}$, and ($\psi_{A-B}$,$\psi_D$) the right actions $$\psi_{A-B}:(\C^*\times S^1)\times S^1\to \C^*\times S^1, \ \psi_{A-B}((z,e^{i\alpha}),e^{i\beta})=(z,e^{i(\alpha+\beta)})\eqno{(22)}$$ and $$\psi_D:S^3\times S^1\to S^3, \ \psi_D((z_1,z_2),e^{i\lambda})=(z_1e^{i\lambda},z_2e^{i\lambda})\eqno{(23)}$$ is the unique {\it bundle morphism} $$\xi_{A-B}\to \xi_D \eqno{(24)}$$ induced by the inclusion $\iota$ i.e. $$\pi\circ\bar{\iota}=\iota\circ pr_1\eqno{(25)}$$ and $$\psi_D\circ(\bar{\iota}\times Id_{S^1})=\bar{\iota}\circ\psi_{A-B}\eqno{(26)}$$ namely, with lower and upper parts of Diagram 1 commuting. $$\matrix{(\C^*\times S^1)\times S^1 & \buildrel{\bar{\iota}\times Id_{S^1}}\over\longrightarrow & S^3\times S^1\cr \psi_{A-B}\downarrow & & \downarrow\psi_D\cr \C^*\times S^1 & \buildrel{\bar{\iota}}\over\longrightarrow & S^3\cr pr_1\downarrow & & \downarrow\pi\cr \C^* & \buildrel{\iota}\over\longrightarrow & \C\cup\{\infty\}\cr}$$ \centerline{{\it Diagram 1}}

\

In fact:

\

$$\pi\circ\bar{\iota}(z,e^{i\varphi})=\pi({{(z,1)}\over{\vert\vert(z,1)\vert\vert}}e^{i\varphi})=z,$$ $$\iota\circ pr_1(z,e^{i\varphi})=\iota(z)=z;$$ $$\psi_D\circ(\bar{\iota}\times Id_{S^1})((z,e^{i\varphi}),e^{i\lambda})=\psi_D(\bar{\iota}(z,e^{i\varphi}),e^{i\lambda})={{(z,1)}\over{\vert\vert(z,1)\vert\vert}}e^{i(\varphi+\lambda)},$$ $$ \bar{\iota}\circ\psi_{A-B}((z,e^{i\varphi}),e^{i\lambda})=\bar{\iota}(z,e^{i(\varphi+\lambda)})={{(z,1)}\over{\vert\vert(z,1)\vert\vert}}e^{i(\varphi+\lambda)}.$$

\

{\bf 4. Pull-back of $\xi_D$ by $\iota$: $\iota^*(\xi_D)$}

\

The total space of the {\it induced} or {\it pull-back} bundle [14] of $\xi_D$ by $\iota$, $\iota^*(\xi_D):S^1\to P_{\iota^*(\xi_D)}\buildrel{pr_1}\over\longrightarrow\C^*$, is defined by $$P_{\iota^*(\xi_D)}=\{(z,(z_1,z_2))\in\C^*\times S^3, \ \iota(z)=\pi(z_1,z_2)\}\eqno{(27)}$$ and must be such that both the upper and lower parts of Diagram 2 commute i.e. such that ($\iota$,$pr_2$) is a bundle morphism $\iota^*(\xi_D)\to\xi_D$. In Diagram 2, $pr_2$ is the projection in the second entry, and $$\psi_{\iota^*(\xi_D)}:P_{\iota^*(\xi_D)}\times S^1\to P_{\iota^*(\xi_D)}, \ \psi_{\iota^*(\xi_D)}((z,(z_1,z_2)),e^{i\lambda})=(z,(z_1,z_2)e^{i\lambda})\eqno{(28)}$$ is the right action of $S^1$ on $P_{\iota^*(\xi_D)}$. $$\matrix{P_{\iota^*(\xi_D)}\times S^1 & \buildrel{pr_2\times Id_{S^1}}\over\longrightarrow & S^3\times S^1\cr \psi_{\iota^*(\xi_D)}\downarrow & & \downarrow\psi_D\cr P_{\iota^*(\xi_D)} & \buildrel{pr_2}\over\longrightarrow & S^3\cr pr_1\downarrow & & \downarrow\pi\cr \C^* & \buildrel{\iota}\over\longrightarrow & \C\cup\{\infty\}\cr}$$ \centerline{{\it Diagram 2}} 

\

From $$\iota\circ pr_1=\pi\circ pr_2\eqno{(29)}$$ one has: $$\iota\circ pr_1((z,(z_1,z_2))=\iota(z)=z,$$ $$\pi\circ pr_2((z,(z_1,z_2))=\pi(z_1,z_2)=z_1/z_2,$$ so $z_1=z_2z$ and $\vert\vert(z_1,z_2)\vert\vert=1$ implies $(z_1,z_2)={{(z,1)}\over{\vert\vert(z,1)\vert\vert}}e^{i\varphi}$. Then, $$P_{\iota^*(\xi_D)}=\{(z,{{(z,1)}\over{\vert\vert(z,1)\vert\vert}}e^{i\varphi}), \ z\in\C^*, \ \varphi\in[0,2\pi)\}\subset\C^*\times S^3.\eqno{(30)}$$ On the other hand, it holds $$\psi_D\circ(pr_2\times Id_{S^1})=pr_2\circ\psi_{\iota^*(\xi_D)}.\eqno{(31)}$$ In fact: $$\psi_D\circ(pr_2\times Id_{S^1})((z,(z_1,z_2)),e^{i\lambda})=\psi_D((z_1,z_2)e^{i\lambda})=
(z_1e^{i\lambda},z_2e^{i\lambda}),$$ $$pr_2\circ\psi_{\iota^*(\xi_D)}((z,(z_1,z_2)),e^{i\lambda})=pr_2((z,(z_1,z_2)e^{i\lambda}))=(z_1,z_2)e^{i\lambda}=(z_1e^{i\lambda},z_2e^{i\lambda}).$$

\

{\bf 5. Bundle isomorphism: $\iota^*(\xi_D)\buildrel{\cong}\over\longrightarrow \xi_{A-B}$}

\

In this section we exhibit a ``natural" isomorphism between the $A-B$ bundle and the pull-back by the inclusion $\iota:\C^*\to \C\cup\{\infty\}$ (i.e. $\iota:(D_0^2)^*\to S^2$ up to homeomorphisms) of the Dirac bundle $\xi_D$ corresponding to unit magnetic charge, thus establishing a deep relation between the two systems ($A-B$: experimentally observed, and $D$: only theoretical, up to now). 

\

The homeomorphism between the total spaces of the bundles is given by $$\Psi:P_{\iota^*(\xi_D)}\to\C^*\times S^1, \ \Psi(z,{{(z,1)}\over{\vert\vert(z,1)\vert\vert}}e^{i\varphi})=(z,e^{i\varphi}).\eqno{(32)}$$ It is clear that $\Psi$ is continuous, one-to-one and onto, with continuous inverse $\Psi^{-1}$. It is easily verified that Diagram 3, corresponding to this isomorphism, commutes in its upper and lower parts i.e. $$pr_1\circ\Psi=Id_{\C^*}\circ pr_1\eqno{(33)}$$ and $$\psi_{A-B}\circ(\Psi\times Id_{S^1})=\Psi\circ\psi_{\iota^*(\xi_D)}.\eqno{(34)}$$ 

$$\matrix{P_{\iota^*(\xi_D)}\times S^1 & \buildrel{\Psi\times Id_{S^1}}\over\longrightarrow & (\C^*\times S^1)\times S^1\cr \psi_{\iota^*(\xi_D)}\downarrow & & \downarrow\psi_{A-B}\cr P_{\iota^*(\xi_D)} & \buildrel{\Psi}\over\longrightarrow & \C^*\times S^1\cr pr_1\downarrow & & \downarrow pr_1\cr \C^* & \buildrel{Id_{\C^*}}\over\longrightarrow & \C^*\cr}$$ \centerline{{\it Diagram 3}} 

\

In fact: $$pr_1\circ\Psi(z,{{(z,1)}\over{\vert\vert(z,1)\vert\vert}}e^{i\varphi})=pr_1(z,e^{i\varphi})=z,$$ $$Id_{\C^*}\circ pr_1(z,{{(z,1)}\over{\vert\vert(z,1)\vert\vert}}e^{i\varphi})=Id_{\C^*}(z)=z;$$ $$\psi_{A-B}\circ(\Psi\times Id_{S^1})((z,(z_1,z_2)),e^{i\lambda})=\psi_{A-B}(\Psi((z,(z_1,z_2)),e^{i\lambda}))=\Psi(z,(z_1,z_2))e^{i\lambda}=\Psi(z,{{(z,1)}\over{\vert\vert(z,1)\vert\vert}}e^{i\varphi})e^{i\lambda}$$ $$=(z,e^{i\varphi})e^{i\lambda}=(z,e^{i(\varphi+\lambda)}),$$ $$\Psi\circ\psi_{\iota^*(\xi_D)}((z,(z_1,z_2)),e^{i\lambda})=\Psi(z,(z_1,z_2)e^{i\lambda})=\Psi(z,{{(z,1)}\over{\vert\vert(z,1)\vert\vert}}e^{i(\varphi+\lambda)})=(z,e^{i(\varphi+\lambda)}).$$

\

{\bf 6. Chern classes}

\

$\xi_{A-B}$ is the pull-back of $\xi_D$ by the inclusion $\iota:(D_0^2)^*\to S^2$; however, since $\xi_{A-B}$ is trivial, then all its Chern classes must vanish. Then, in particular, we must verify the vanishing of the pull-back of $c_1$.
\

$\xi_{A-B}=\iota^*(\xi_D)$ passes to cohomology [14] in the form $$\iota^*:H^*(S^2)\to H^*((D_0^2)^*)\eqno{(35a)}$$ i.e. $$\iota^*:H^k(S^2)\to H^k((D_0^2)^*), \ k=0,1,2\eqno{(35b)}$$ where $$H^*(S^2)=(H^0(S^2),H^1(S^2),H^2(S^2))\cong(\R,0,\R)\eqno{(36)}$$ and $$H^*((D_0^2)^*)=(H^0((D_0^2)^*),H^1((D_0^2)^*),H^2((D_0^2)^*))\cong(\R,\R,0)\eqno{(37)}$$ are the cohomology groups of the 2-sphere and the punctured disk respectively. $H^*((D_0^2)^*)\cong H^*(S^1)$ by homotopy invariance. Since $c_1\in H^2(S^2)$, then $$\iota^*(c_1)=0.\eqno{(38)}$$

\

{\bf 7. Pull-back of the Dirac connection and vanishing of the $A-B$ effect}

\

In terms of the cartesian coordinates in $\R^3$, $(x,y,z)=r(sin\theta cos\varphi,sin\theta sin\varphi,cos\theta)$ with $\theta\in(0,\pi)$ and $\varphi\in [0,2\pi)$ which implies $(x,y,z)\neq (0,0,z)$, the monopole potentials $A_\pm$ of equations (16a) and (16b) are given by $$A_\pm(x,y,z)=(A_\pm)_xdx+(A_\pm)_ydy\eqno{(39)}$$ with $$(A_\pm)_x(x,y,z)=\pm {{i}\over{2}}({{y}\over{x^2+y^2}})(1\mp{{z}\over{\sqrt{x^2+y^2+z^2}}}), \ (A_\pm)_y(x,y,z)=\mp{{i}\over{2}}({{x}\over{x^2+y^2}})(1\mp{{z}\over{\sqrt{x^2+y^2+z^2}}}).\eqno{(40)}$$ (Notice that $[(A_\pm)_x]=[(A_\pm)_y]=[L]^{-1}$ since $[x]=[y]=[z]=[L]$ while $[A_\pm]=[L]^0$, $L$: length.)

\

To pull-back by $\iota$ these 1-forms to $(D_0^2)^*$ we must first restrict $A_\pm$ to $z=0$ and then perform the pull-back operation, which reduces to the identity: $$\iota^*(A_\pm(x,y,0))=\pm {{i}\over{2}}({{ydx-xdy}\over{x^2+y^2}}):= ia_\pm(x,y)\eqno{(41)}$$ with $$a_\pm(x,y)=\mp{{1}\over{2}}({{xdy-ydx}\over{x^2+y^2}})\eqno{(42)}$$ the real-valued $A-B$ potential 1-forms. Clearly, $a_\pm$ are closed ($da_\pm=0$) but not exact since $a_\pm=\mp{{1}\over{2}}d\varphi$ only for $\varphi\in (0,2\pi)$. If we surround the thin solenoid in the $A-B$ side with closed curves $\gamma_\pm$ with $\gamma_-=-\gamma_+$, then the surrounded magnetic flux is $$\Phi_{A-B}=\int_{\gamma_+}a_++\int_{\gamma_-}a_-=\int_{\gamma_+}a_++\int_{\gamma_-}(-a_+)=\int_{\gamma_+}a_+-\int_{\gamma_+}(-a_+)=2\int_{\gamma_+}a_+=2\int_{\gamma_+}(-{{1}\over{2}}d\varphi)=-2\pi,\eqno{(43)}$$ which coincides, up to a sign, with the flux of the monopole: $$\Phi_D=\int_{S^2}\vec{B}=({{1}\over{2}})\int_{S^2}{{\hat{r}\cdot\hat{r}}\over{r^2}}=({{1}\over{2}})4\pi=2\pi.\eqno{(44)}$$ But this implies that the $A-B$ effect vanishes if and only if the value of the electric charge $\vert e\vert$ is an integer: the $D.Q.C.$ for the present case where $g={{1}\over{2}}$. In fact, with $\Phi_0={{2\pi}\over{\vert e\vert}}$ the quantum of magnetic flux associated with the charge $\vert e\vert$, the phase change of the wave function in the $A-B$ experiment due to the presence of magnetic flux is $$e^{-i\vert e\vert\Phi_{A-B}}=e^{-2\pi i{{\Phi_{A-B}}\over{\Phi_0}}}=e^{2\pi i{{\Phi_D}\over{\Phi_0}}}=e^{i\vert e\vert({{1}\over{2}})4\pi}=e^{2\pi i\vert e\vert}=1\Leftrightarrow \vert e\vert=n\in\Z.\eqno{(45)}$$ (For arbitrary $g$, the $D.Q.C.$ would be $\vert e\vert g={{n}\over{2}}$.)

\

{\bf 8. Final comments}

\

It is well known that the $A-B$ effect and the Dirac monopole are closed related [15]; in particular the disappearance of the Dirac string simultaneously with the vanishing of the $A-B$ effect when appropriate conditions of the magnetic fluxes are fulfilled [16]. In the present paper, the above relation has been described in the context of the fiber bundles associated with both phenomena, respectively $\xi_{A-B}$ (trivial) and $\xi_D$ (non-trivial Hopf bundle). The remarkable fact is that $\xi_{A-B}$ turns out to be the pull-back of $\xi_D$ by the inclusion $\iota$ of the corresponding base spaces, which allows to discuss the above relation in a purely geometric language. It would be interesting to investigate if this bundle theoretic relation exists in non-abelian cases.

\

{\bf Acknowledgments}

\

The author thanks Gregory L. Naber and Sebasti\'an N\'ajera Valencia for useful and enlightening discussions.

\

{\bf References}

\

[1] Aharonov, Y. and Bohm, D. (1959). {\it Significance of electromagnetic potentials in the quantum theory}, Physical Review {\bf 15}, 485-491.

\

[2] Dirac, P.A.M. (1931). {\it Quantised singularities in the Electromagnetic Field}, Proceedings of the Royal Society {\bf A133}, 60-72.

\

[3] Dirac, P.A.M. (1948). {\it The Theory of Magnetic Poles}, Physical Review {\bf 74}, 817-830.

\

[4] Chambers, R.G. (1960). {\it Shift of an electron interference pattern by enclosed magnetic flux}, Physical Review Letters {\bf 5}, 3-5.

\

[5] Peshkin, M. and Tonomura, A. (1989). {\it The Aharonov-Bohm Effect}, Springer, Berlin.

\

[6] Preskill, J. (1984). {\it Magnetic Monopoles}, Annual Review of Nuclear and Particle Science {\bf 34}, 451-530.

\

[7] Polchinski, J. (2004). {\it Monopoles, Duality, and String Theory}, International Journal of Modern Physics A{\bf 19}, 145-154; arXiv:hep-th/0304042v1.

\

[8] Aguilar, M. and Socolovsky, M. (2002). {\it Aharonov-Bohm Effect, Flat Connections and Green's Theorem}, International Journal of Theoretical Physics {\bf 41}, 839-860; Socolovsky, M. (2006). {\it Aharonov-Bohm Effect}, Encyclopedia of Mathematical Physics, eds. J.P. Francoise, G.L. Naber, and T.S. Sun; Elsevier; 191-198.

\

[9] Naber, G.L. (1997). {\it Topology, Geometry, and Gauge Fields. Foundations}, Springer-Verlag, N.Y. 

\

[10] Steenrod, N. (1951). {\it The Topology of Fibre Bundles}, Princeton University Press, N.J.

\

[11] Socolovsky, M. (1992). {\it Spin, Monopole, Instanton and Hopf Bundles}, Aportaciones Matem\'aticas, Notas de Investigaci\'on {\bf 6}, Sociedad Matem\'atica Mexicana, 141-164.

\

[12] Hopf, H. (1931). {\it $\ddot{U}$ber die Abbildungen der 3-Sph$\ddot{a}$re auf die Kugelfl$\ddot{a}$che}, Mathematische Annalen {\bf 104}, 637-665.

\

[13] Trautman, A. (1977). {\it Solutions of the Maxwell and Yang-Mills Equations Associated with Hopf Fibrings}, International Journal of Theoretical Physics {\bf 16}, 561-565.

\

[14] Husemoller, D. et al. (2008). {\it Basic Bundle Theory and K-Cohomology Invariants}, Lecture Notes in Physics {\bf 726}, Springer, Berlin Heidelberg.

\

[15] Wu, T.T. and Yang, C.N. (1975). {\it Concept of nonintegrable phase factors and global formulation of gauge fields}, Physical Review D {\bf 12}, 3845-3857.

\

[16] Rajantie, A. (2012). {\it Introduction to magnetic monopoles}, Contemporary Physics {\bf 53}, 195-211.

\

\

\

\

\

\

\

* E-mail: {\it socolovs@nucleares.unam.mx}

\end

\centerline{\bf FIBRE BUNDLES AND GAUGE TRANSFORMATIONS}

\

{\bf 1. Principal fibre bundle $\xi$}

\

A smooth principal fibre bundle (p.f.b.) $\xi$ consists of a sextet $$\xi=(P,\pi,M,G,\xi,{\cal U})$$ where $P$ and $M$ are $m+s$ and $m$ dimensional differentiable manifolds, $G$ is an $s$ dimensional Lie group, $P\buildrel{\pi}\over\rightarrow M$ is a projection, $P\times G\buildrel{\psi}\over\rightarrow P$ is an action ($pe=p$ and $p(g_1g_2)=(pg_1)g_2$) free over $P$ ($pg=p\Rightarrow g=e$, $e$ the identity in $G$) and transitive on fibers $P_x=\pi^{-1}(\{x\})$ (if $p,q\in P$ then there exists $g\in G$ such that $q=pg$), and ${\cal U}$ is the atlas of the bundle i.e. the set of local trivializations $P_U \buildrel{\phi_U}\over\rightarrow U\times G$ with $U$ open subset of $M$, $P_U=\pi^{-1}(U)$, and $\phi_U$ a diffeomorphism $\phi(p)=(\pi(p),\gamma_U(p))$ with $\gamma_U(pg)=\gamma_U(p)g$. 

\

$\xi$ is denoted by $G^s\to P^{m+s}\buildrel{\pi}\over\rightarrow M^m$.

\

A section of $P$ is a smooth function $M\buildrel{\sigma}\over\rightarrow P$ such that $\pi\circ\sigma=Id_M$.

\

{\it 1.1 Gauge group of} $\xi$: ${\cal G}(\xi)$

\

${\cal G}(\xi)$ is the set of diffeomorphisms $P\buildrel{\alpha}\over\rightarrow P$ such that the following diagram commutes: $$\matrix{P\times G & \buildrel{\alpha\times Id_G}\over\rightarrow & P\times G \cr \psi\downarrow & & \downarrow\psi \cr P & \buildrel{\alpha}\over\rightarrow & P\cr \pi\downarrow & & \downarrow\pi\cr M & \buildrel{Id_M}\over\rightarrow & M\cr}$$ i.e. $\alpha(pg)=\alpha(p)g$ and $\pi(\alpha(p))=\pi(p)$ (so $\alpha(p)\in P_{\pi(p)}$). ${\cal G}(\xi)$ is called the {\it group of gauge transformations} or {\it vertical automorphisms} of $\xi$.

\

{\it 1.2 Local form of} ${\cal G}(\xi)$

\

{\it Proposition}: For each $(U,\phi_U)\in{\cal U}$ there exists a smooth function $$\alpha_U:U\to G, \ x\mapsto\alpha_U(x)$$ which determines $\alpha\in{\cal G}(\xi)$ for all $p$ in $P_U$.

\

{\it Proof}: $\alpha(\sigma_U(x))=\sigma_U(x)\alpha_U(x)$ for some unique $\alpha_U(x)\in G$, where $U\buildrel{\sigma_U}\over\rightarrow P_U$ is the local section of $P$ given by $\sigma_U(x)=\phi_U^{-1}(x,e)\in P_U$. If $p\in P_x$, then $p=\sigma_U(x)h$ for some unique $h\in G$; so $\alpha(p)=\alpha(\sigma_U(x)h)=\alpha(\sigma_U(x))h=(\sigma_U(x)\alpha_U(x))h=\sigma_U(x)(\alpha_U(x)h)$ i.e. $\alpha_U$ gives $\alpha$ for all $p\in P_x$ for all $x\in U$. qed

\

{\it Corollary}: If $\xi$ is trivial i.e. if $P\cong M\times G$, then $U=M$ and there exists $\alpha_M:M\to G$. qed

\

Clearly, if $\xi$ is trivial, then $M\buildrel{\sigma}\over\rightarrow P$ with $\sigma(x)=\phi^{-1}(x,e)$ is a section of $P$, where $\phi$ is the diffeomorphism $P\buildrel{\phi}\over\rightarrow M\times G$. The other way around: if $\xi$ has a section $M\buildrel{\sigma}\over\rightarrow P$, $\sigma(x)=p$, then $P\buildrel{\phi}\over\rightarrow M\times G$ with $\phi(q)=(\pi(q),g), \ q=pg$, is a trivialization of $P$. So, a p.f.b. is trivial if and only if it has a (global) section.  

\

{\it 1.3 Example: Frame bundle of a differentiable manifold}

\

$$\xi_{\cal F}: \ \ (GL_m(\R))^{m^2}\to (FM^m)^{m^2+m}\buildrel{\pi_{\cal F}}\over\rightarrow M^m$$ where $$FM^m=\bigcup_{x\in M^m}\{x\}\times\{(v_{1x},\cdots,v_{mx})\},$$ with $$(v_{1x},\cdots,v_{mx})\equiv r_x$$ an ordered basis ({\it Vielbeine}) of $T_xM^m$, $FM^m\times GL_m(\R)\buildrel{\psi}\over\rightarrow FM^m, \ \psi(r_x,a)=(v_{\nu x}a^\nu_1,\cdots,v_{\nu x}a^\nu_m)$, local coordinates on $FM^m$ given by $x^\rho(x,r_x)=x^\rho(x)$ and $X^\mu_\nu(x,r_x)=v^\mu_{\nu x}$, and $\pi_{\cal F}(x,r_x)=x$. If $\{{{\partial}\over{\partial x^\mu}}\}_{\mu=1}^m$ is a local coordinate basis of $\Gamma(TU)$, then $v_{kx}=v_ k^\mu(x){{\partial}\over{\partial x^\mu}}\vert_x$, with $x\in U$, $k=1,\dots,m$.

\

In $\xi_{\cal F}$ there exists the {\it canonical form} $\theta:FM^m\to T^*FM^m\otimes\R^m$, $\theta((x,r_x))=((x,r_x),\theta_{(x,r_x)})$, and $\theta_{(x,r_x)}:T_{(x,r_x)}FM^m\to\R^m$ given by the composition $\tilde{r}_x\circ d\pi_{\cal F}\vert_{(x,r_x)}$ where $\tilde{r}_x:T_xM^m\to \R^m$, $\tilde{r}_x(\lambda^\nu v_{\nu x})=(\lambda^1,\dots,\lambda^m)$. So, $\theta$ solders the tangent spaces of $M^m$ to the tangent spaces of $FM^m$. 

\

{\bf 2. Associated bundle} $\xi_F$

Let $\xi$ be a p.f.b., $F$ an $n$ dimensional differentiable manifold, and $G\times F\buildrel{\mu}\over\rightarrow F$ a left action of $G$ on $F$. Then $\xi$ and $\mu$ induce the bundle $$\xi_F=(P_F,\pi_F,M,F,\mu,{\cal U}_F)$$ with total space the $n+m$ dimensional differentiable manifold $$P_F=P\times_GF=\{[(p,f)]\}_{(p,f)\in P\times F}, \ [(p,f)]=\{(pg,g^{-1}f)\}_{g\in G},$$ fiber $F$, projection $$P_F\buildrel{\pi_F}\over\rightarrow M, \ \pi_F([(p,f)])=\pi(p),$$ and atlas ${\cal U}_F$ given by the local trivializations $$P^F_U\buildrel{\phi^F_U}\over\rightarrow U\times F, \ \phi^F_U([(p,f)])=(\pi(p),\gamma_U(p)f),$$ where $P^F_U=\pi_F^{-1}(U)$. 

\

($\phi^F_U$ is well defined since $\phi^F_U([(pg,g^{-1}f)]=(\pi(pg),\gamma_U(pg)g^{-1}f)=(\pi(p),\gamma_U(p)gg^{-1}f)=\phi^F_U([(p,f)]$.)

\

$\xi_F$ is called an {\it associated bundle} (a.b.) and is denoted by $F^n-P_F^{n+m}\buildrel{\pi_F}\over\rightarrow M^m$.

\

The space of {\it sections} of $\xi_F$ is the set of smooth functions $$\Gamma(P_F)=\{s:M\to P_F \ \vert \ \pi_F\circ s=Id_M\}.$$

\

The space of {\it equivariant functions} $P-F$ is the set of smooth functions $$\Gamma_{eq.}(P,F)=\{\gamma:P\to F \ \vert \ \gamma(pg)=g^{-1}\gamma(p)\}.$$

\

{\it Proposition}: There is a one-to-one correspondence between $\Gamma(P_F)$ and $\Gamma_{eq.}(P,F)$.

\

{\it Proof}: Given $s\in \Gamma(P_F)$, there exists $\gamma_s:P\to F$, $\gamma_s(p)=f$ where $s(\pi(p))=[(p,f)]$. Given $\gamma$, there exists $s_\gamma:M\to P_F$, $s_\gamma(x)=[(p,f)]$ with $\gamma(p)=f$ and $p\in\pi^{-1}(\{x\})$. qed

\

This is summarized in the following diagram:

\

$$\matrix{G & & F\cr \downarrow & \gamma\nearrow & \vert\cr P & & P_F\cr \pi\downarrow & & s\uparrow\downarrow\pi_F\cr M & & M\cr}$$

\

If $F$ is a vector space $V$ and $\mu$ is a linear representation of $G$ over $V$, then $\xi_V$ is a {\it vector bundle} (v.b.) associated to $\xi$. Typically $V=\R^n$ or $\C^n$. If $\{e_1,\cdots,e_n\}$ is a basis of $V$, then $$U\buildrel{\sigma^V_{U,k}}\over\rightarrow P^V_U, \ \sigma^V_{U,k}(x)=[(p,(\gamma_U(p))^{-1}e_k)], \ p\in P_x,$$ $k=1,\cdots,n$, is a set of $n$ linearly independent local sections of $\xi_V$. Typically, $V$ is the representation space of an irreducible representation of $G$.

\

{\it 2.1 Gauge transformations of sections}

\

Let $\alpha\in{\cal G(\xi)}$ and $s\in\Gamma(P_F)$, then $\alpha^*(\gamma_s)=\gamma_s\circ\alpha\in\Gamma_{eq.}(P,F)$ since $\gamma_s\circ\alpha(pg)=\gamma_s(\alpha(pg))=\gamma_s(\alpha(p)g)=g^{-1}\gamma_s(\alpha(p))=g^{-1}\gamma_s\circ\alpha(p)=g^{-1}\alpha^*(\gamma_s)(p)$. Then $$\tilde{\alpha}(s)=s_{\alpha^*(\gamma_s)}$$ is the gauge transformed of the section $s$, with $$\tilde{\alpha}(s)(x)=[(p,g^{-1}f)]$$ where $s(x)=[(p,f)]$, $\alpha(p)=pg$, and $p\in\pi^{-1}(\{x\})$.

\

{\it 2.2 Example: Bundle of tensors $T^r_sM^m$ over a differentiable manifold}

\

$G=GL_m(\R)$, $F=V=\R^{m^{r+s}}$, $r,s\in\{0,1,2,3,\dots\}$, $$\mu(a,\vec{\lambda})_{j_1,\dots,j_s}^{i_1,\dots,i_r}=a^{i_1}_{k_1}\dots a^{i_r}_{k_r}(a^{-1})^{l_1}_{j_1}\dots (a^{-1})^{l_s}_{j_s}\lambda^{k_1\dots k_r}_{l_1\dots l_s}.$$ One has the bundle isomorphism $\varphi$ given by the diagram: 

$$\matrix{\R^{m^{r+s}} & & \R^{m^{r+s}} \cr \vert & & \vert \cr FM^m\times_{GL_m(\R)}\R^{m^{r+s}} & \buildrel{\varphi}\over\longrightarrow & T^r_sM^m \cr \pi_V\downarrow & &  \downarrow\pi^r_s \cr M^m & \buildrel{Id_{M^m}}\over\longrightarrow & M^m \cr},$$ where
$\varphi([(r_x,\vec{\lambda})])=\sum^m_{i_k,j_l=1}\lambda^{i_1\dots i_r}_{j_1\dots j_s}v_{i_1x}\otimes\dots v_{i_rx}\otimes w_x^{j_1}
\otimes\dots\otimes w_x^{j_s}$, with $\{w_x^i\}$ the dual basis of $\{v_{jx}\}$: $w_x^i(v_{jx})=\delta^i_j.$

\

{\bf 3. Connection in a principal fibre bundle}

\

{\it 3.1 Gauge transformations of connections}

\

{\bf 4. Covariant derivative of sections}

\

{\it 4.1 Gauge transformations of covariant derivative}

\

{\it 4.2 Local form of covariant derivative}

\

{\it 4.3 Parallel transport}

\

\

\

\

\

\

\

\

\

e-mail: socolovs@nucleares.unam.mx

\end